\documentclass[12pt]{article}

\usepackage{graphicx}
\usepackage{amsmath}
\usepackage{array}
\usepackage{setspace}

\newcommand{\piiofx}{\pi_{i}(\mathbf{x})}
\newcommand{\piNofx}{\pi_{M}(\mathbf{x})}
\newcommand{\piNminusone}{\pi_{M-1}(\mathbf{x})}

\newcommand{\intApitwo}{\int d\mathbf{x}\, A(\mathbf{x})\pi_{M-1}(\mathbf{x})}
\newcommand{\intApionew}{\int d\mathbf{x}\, A(\mathbf{x})\pi_{M}(\mathbf{x})w(\mathbf{x})}
\newcommand{\Zint}{\int d\mathbf{x}\, \piiofx}
\newcommand{\what}{w(\mathbf{x}_{j}(t_{0}))}

\newcommand{\xjoft}{\mathbf{x}_{j}(t_{i-1})}

\title{Annealed importance sampling of dileucine peptide}
\author{Edward Lyman\footnote{elyman@ccbb.pitt.edu},
and Daniel M. Zuckerman\footnote{dmz@ccbb.pitt.edu}\\
    Department of Computational Biology, School of Medicine\\
3079 BST3 , 3501 Fifth Ave., \\University
of Pittsburgh, Pittsburgh, PA 15261}

\begin{document}
\maketitle

\abstract{Annealed importance sampling is a means to assign equilibrium weights to a nonequilibrium 
 sample that was generated by a simulated annealing protocol\cite{neal-ais01}. The weights may then be used to calculate 
 equilibrium averages, and also serve as an ``adiabatic signature'' of the chosen cooling schedule. 
 In this paper we demonstrate the method 
 on the $50$-atom dileucine peptide, showing that equilibrium distributions are attained for manageable cooling schedules. For this 
 system, as na\"ively 
 implemented here, the method is modestly more efficient 
 than constant temperature simulation. However, the method is worth considering whenever any 
 simulated heating or cooling is performed (as is often done at the beginning of a simulation project, 
 or during an NMR structure calculation), as it is simple to implement and requires minimal additional CPU expense. 
 Furthermore, the na\"ive implementation presented here can be improved.}
\section{Introduction}
Simulated annealing (SA) is used in a wide variety of biomolecular calculations. 
Crystallographic refinement protocols\cite{brunger-cns} and standard NMR structure 
calculations\cite{xplor-nih,wutrich-dyana-jmb97,brunger-cartSA-structure97} 
both rely on SA to optimize a ``target function,'' constructed so that the global minimum of 
the target function corresponds to the native structure. Molecular dynamics calculations often 
begin by cooling a configuration from a high temperature ensemble to a lower temperature, 
at which the simulation is to be performed. 

In this paper, we consider a different use for SA calculations. Since a set of structures that is 
generated by a series of SA trajectories is a nonequilibrium sample, they may not be used 
to calculate equilibrium averages. However, Neal demonstrated a simple procedure, 
called ``annealed importance sampling'' (AIS)
that allows the nonequilibrium sample to be reweighted into an equilibrium one\cite{neal-ais01}. AIS 
is closely connected with the Jarzynski relation\cite{jarzynski}. To 
our knowledge, the algorithm has only appeared once in the chemical physics literature\cite{coolwalk}, 
where it was used (along with sophisticated Monte Carlo techniques) to sample a one-dimensional 
potential. Here, we demonstrate an application of the AIS algorithm to generate an equilibrium sample 
of an implicitly solvated peptide, and discuss other uses for AIS which may of interest to the molecular 
simulation community.

The basic idea which underlies SA is also the motivation for other temperature based sampling methods, 
notably J-walking\cite{jwalk-jcp90}, simulated tempering\cite{simtemp-jcp92,simtemp-ephyslett92} and 
replica exchange/parallel tempering\cite{Geyer-1991,deem-rev-pccp05}. 
By coupling a simulation to a high temperature reservoir, it is hoped that the low temperature simulation 
may explore the configuration space more thoroughly. This is achieved by thermally activated crossing of energetic barriers, 
which are large compared to the thermal energy scale of the lower temperature simulation, but are crossed 
more frequently at higher temperature. Simulated and parallel tempering differ in the way that the 
different temperature simulations are coupled. Simulated tempering heats and then cools the system, in a way that 
maintains an equilibrium distribution. Parallel tempering couples simulations run in parallel at different 
temperatures by occasionally swapping configuartions between temperatures, again in such a way that canonical 
sampling is maintained. 

AIS offers yet another approach to utilizing a high temperature ensemble for equilibrium sampling at a lower 
temperature. A sample of a high temperature ensemble is annealed to a lower temperature, by alternating 
constant temperature simulation with steps in which the tempertaure is jumped to a lower value. Each annealed 
structure is assigned a weight, which depends on the trajectory that was traced during the annealing process. 
Equilibrium averages over the lower temperature ensemble may then be calculated by a simple weighted average. 
Furthermore, the distribution of trajectory weights contains useful information about the statistics of the 
annealed sample. Roughly, a schedule which quenches high temperature structures very rapidly to low temperature 
will result in a sample dominated by a few high weight structures, resulting in poor statistics. This connection 
between the distribution of weights and the extent to which the schedule is not adiabatic ought to be of 
interest to anyone who uses SA protocols---whether for equilibrium sampling or for structure calculation.

We have used the AIS method to generate $298$ K equilibrium ensembles of the dileucine peptide, 
by annealing structures from a $500$ K distribution with several different cooling schedules. 
For the most efficient schedule used, we found a modest gain (about a factor of $3$) over 
constant temperature simulation. This result is consistent with earlier observations on the expected efficiency 
of temperature-based sampling methods\cite{repex-note}.

\section{Theory}

Consider a standard simulated annealing (SA)
trajectory, in which a protein is slowly cooled from a conformation $\mathbf{x}$ at a (high)
temperature $T_{M}$. The cooling is achieved by alternating constant temperature dynamics 
with ``temperature jumps,'' during which the temperature is lowered instantaneously. 
Usually, the system is cooled to a low temperature, since the
aim of standard SA calculations is to find the global minimum on the energy landscape. 
But we can imagine instead ending the run
at $T_{0}=300$ K---in fact, we can think of many such runs, all ending
at $300$ K. We then have an ensemble of conformations, though clearly not distributed
canonically at $T_{0}$. We would like to know if there is a way to \emph{reweight} this
distribution, so that it can be used to compute equilibrium averages at $T_{0}$. The
affirmative answer is provided by the annealed importance sampling (AIS) method.

To make the discussion more concrete, consider many independent annealing trajectories
$\mathbf{x}_{j}(t)$ which at time $t_{M-1}$ have just been
cooled from inverse temperature $\beta_{M}$ to $\beta_{M-1}$. As usual, each temperature
defines
a distribution of conformations: $\piiofx \propto \exp[-\beta_{i}U(\mathbf{x})]$. 
Immediately after $t_{M-1}$, before the system is allowed
to relax to $\piNminusone$, we can compute the equilibrium average of an arbitrary
quantity $A$
over $\piNminusone$ by using the weight $w(\mathbf{x})=\piNminusone/\piNofx$:
\begin{equation}
\langle A\rangle _{M-1}Z_{M-1} = \intApitwo = \intApionew,
\label{singleweight}
\end{equation}
where $\langle A\rangle_{i}$ denotes an average over $\pi_{i}$, and $Z_{i}=\Zint$.
In other words, we may reweight the distribution $\pi_M(\mathbf{x})$ to calculate
averages over $\pi_{M-1}(\mathbf{x})$, by multiplying by the ratio of Boltzmann factors.

Generalizing the argument to $M$ temperature steps is straightforward\cite{neal-ais01}, by
forming the product of weights for successive cooling steps:
\begin{equation}
w_j \equiv \what = \prod^{M}_{i=1}\frac{\pi_{i-1}(\xjoft)}{\pi_{i}(\xjoft)}.
\label{weightprod}
\end{equation}
Equation \ref{weightprod} gives the weight for trajectory $j$, cooled at successive
times $t_{M-1}$, $t_{M-2}$,... through inverse temperatures $\beta_{M}$,
$\beta_{M-1}$,... to reach conformation $\mathbf{x}_{j}(t_{0})$.
At each temperature, reweighting ensures that averages may be calculated for the
appropriate canonical distribution, even though the system has not yet relaxed.

The AIS idea is easily turned into an algorithm for producing a canonical
distribution from serially generated annealing trajectories:

\parbox{4in}{
        (i) Generate a sample of the distribution $\piNofx$, by a sufficiently long
	    simulation at $T_{M}$.\\
        (ii) Pull a conformation from $\piNofx$ at random and anneal down to $\beta_{0}$,
        yielding conformation $\mathbf{x}_1(t_{0})$.
        Keep track of the weight $w(\mathbf{x}_1(t_{0}))$ for this trajectory
	    by Eq.~\ref{weightprod}.\\
	(iii) Repeat steps (iii) and (iv) $N$ times, yielding congiurations $\mathbf{x}_j$ and weights 
	$w(\mathbf{x}_j)\equiv w_j$ for $j=1,1,...,N$.\\
}

Equilibrium averages at temperature $T_0$ are then calculated by a weighted average:
\begin{equation}
\langle A\rangle _0 = \frac{\sum_{j=1}^N w_j A_j}{\sum_{j=1}^N w_j}
\label{weightedavg}
\end{equation}

The cooling schedule is defined by the number and spacing of the temperature steps, as well as
the duration of the constant temperature simulation at each step. As available resources necessarily
limit the CPU time spent on each annealing trajectory, careful consideration of the schedule is in order.
Clearly, a schedule in which high temperature configurations are quenched in one step to low temperature
amounts to a single-step reweighting procedure\cite{swendsen-singhist}. We may expect that such a schedule
would be quite ineffective for large temperature jumps, since very few configurations in the high temperature
distribution have appreciable weight in the low temperature distribution. By introducing intermediate
steps, the system is allowed to relax locally, bridging the high and low temperature distributions in
a way that echoes replica exchange protocols\cite{Geyer-1991,deem-rev-pccp05}, 
simulated tempering\cite{simtemp-jcp92,simtemp-ephyslett92},
and the multiple histogram method\cite{swendsen-wham}. However, the ``top-down'' structure of the 
algorithm most closely resembles J-walking\cite{jwalk-jcp90,jwalk-multistage}.

\section{Results}
The dileucine peptide (ACE-[Leu]$_2$-NME) is good choice for the validation of new algorithms, as 
it is small enough ($50$ atoms, including nonpolar hydrogens) that exhaustive sampling by standard 
simulation methods is possible, yet more akin to protein systems than a one- or two-dimensional 
``toy'' model. 

The high temperature ensemble was generated by $300$ nsec of Langevin dynamics at $T_M = 500$ K, as implemented in 
Tinker v. $4.2.2$\cite{tinker-web}, with a timestep of $1.0$ fsec, and a friction constant of $91$ psec$^{-1}$, 
and solvation was treated by the GB/SA method\cite{still-gbsa}. Frames were written every psec, resulting in 
a sample of $3\times10^4$ frames in the high temperature sample. 

The $500$ K sample was annealed down to $298$ K using $4$ different schedules, consisting of a total of $3$, $5$, $9$, 
and $17$ temperatures, including the endpoints. In each case, the temperatures were distributed geometrically. 
Following each 
temperature jump, the velocities were reinitialized by sampling randomly from the Maxwell-Boltzmann distribution, 
and then allowed to relax at constant temperature for a time $t_R = 0.5$ psec (except where noted) with the protocol described
above. A total 
of $N=1.6\times10^4$ annealing trajectories were generated for each schedule. The control of 
the integration routine to effect the annealing, as well as the calculation of the trajectory weights, were implemented 
in a Perl script.

Figure \ref{ehistofig} shows that the $298$ K distribution of energy is recovered by the AIS 
procedure. It is noteworthy that the $500$ K distribution (corresponding to the high $T$ sample)  
overlaps very little with the $298$ K distribution, and yet the $298$ K distribution is reproduced well for the  
two slowest schedules. Equally interesting is how poorly the algorithm performs when the structures are cooled 
too rapidly, especially on the low $E$ side of the distribution, where there is no overlap with the high $T$ 
distribution. We conclude that the schedules with $3$ or $5$ $T$-steps quench the structures too rapidly, 
resulting in many of the trajectories becoming ``stuck'' in high-energy states that are metastable at $298$ K.

This last observation may be quantified by asking, ``How many of the annealed structures contribute appreciable 
weight to averages calculated with Eq.\ \ref{weightedavg}?'' To address this question, for each schedule we estimated the 
number of configurations $n$ which contribute appreciable weight to the averages: 
\begin{equation}
n \equiv \frac{\sum_{i=1}^N w_i}{w_{\text{max}}}\equiv fN,
\label{ndef}
\end{equation}
where $w_{max}$ is the largest weight observed (see Table \ref{table1}). 
If this number is near $1$, then a small number of trajectories dominate the average---see Eq.\ \ref{weightedavg}
---and poor results should be expected. The effective fraction of the annealing trajectories which generate ``useful'' or 
``successful'' structures is denoteed by $f$.

A more complete picture is provided by the full distribution of the (logarithm of) trajectory weights 
(Fig.\ \ref{weighthistofig}). For each schedule, the 
weights which contribute the most to the $T = 298$ K sample are to the right, at large values of $w$. 
The trend is clear---as slower cooling is effected, the distribution narrows and shifts to the right. It has been 
shown that the accuracy of averages computed from this type of protocol is roughly related to the variance of 
the (adjusted) weights\cite{neal-ais01}. (The adjusted weight is the weight divided by the average weight.) 
This ``rule of thumb'' is borne out by the data in Fig.\ \ref{weighthistofig} and Table \ref{table1}---as the 
cooling slows down the distribution of weights narrows, and the number of trajectories contributing to the 
equilibrium averages increases. This type of analysis may serve
as a means of distinguishing between annealing schedules to decide on a cooling schedule which is slow enough
to yield reasonable estimates of equilibrium averages. It is also essential for optimizing an 
AIS protocol for sampling efficiency, as discussed in the next few paragraphs.

How much better than standard simulation (if at all) is equilibrium sampling by AIS? 
In order to make a direct comparison between AIS and constant temperature simulation, we need to compare the 
CPU time invested per \emph{statistically independent configuration} in each protocol. For the constant temperature simulation, 
this time may be estimated in several ways\cite{ferrenberg,decorr-analysis}, and is essentially the time needed for 
the simulation to ``forget'' where it has been. Following the convention for correlation times, we call this 
time $\tau_i = \tau(T_i)$, where $i$ labels the temperature: $M$ for the high $T$ distribution, and $0$ for the low $T$ 
distribution. For the system studied here, $\tau_M=0.8$ nsec and $\tau_0=3.0$ nsec, as estimated from timseries 
of the $\alpha\rightarrow\beta$ backbone dihedral transition\cite{repex-note}.

The total cost to generate  a structure in an AIS simulation is the sum of the costs 
of generating a structure in the high $T$ distribution plus that for the annealing phase. 
Of course, not every annealing trajectory contributes to thermodynamic averages(Eq.\ \ref{weightedavg}).  
What then is the total cost $t_{\text{cost}}$ of a ``successful'' annealed structure? 
The first part is from high temperature sampling---i.e., $\tau_M$. The second part is the cost of 
all the annealing trajectories, divided by the number which contribute to equilibrium averages.
The time $t_{\text{anneal}}$ is the time spent annealing each structure:
\begin{equation}
t_{\text{anneal}} = t_R(M-2)
\label{t-anneal}
\end{equation}
Recall that $t_R$ is the duration of the constant temperature relaxation steps, and there is no relaxation phase at the
highest and lowest temperatures.

The total cost $t_{\text{cost}}$ is then the sum of $\tau_M$ and $t_{\text{anneal}}$: 
\begin{equation}
t_c = \tau_M + t_{\text{anneal}}/f.
\label{aiscost}
\end{equation}
The efficiency of an AIS protocol may then be computed by taking the ratio 
$R \equiv \tau_0/t_{\text{cost}}$ (see Table \ref{table1}), which gives the factor by which an AIS protocol 
is more or less efficient than constant temperature simulation.
The data in Table \ref{table1} show that the best schedule used here offer a modest speedup over constant 
temperature simulation, of a factor of about $3$. These findings are in agreement with an analysis we have 
published of another temperature-based sampling protocol\cite{repex-note}. We note that an optimized AIS 
protocol would require tuning $N$ based on (perhaps preliminary) estimates of $f$.

It is instructive to compare the AIS results to simple reweighting---i.e., AIS with no intermediate temperature 
steps or relaxation. In this case, no computer time is spent annealing, and the efficiency gain is simply 
$\tau_0/\tau_M = 3.75$. The fraction $f$ is of course reduced compared to any AIS protocol---when reweighting 
our $500$ K dileucine trajectory to $298$ K distribution, $f=1.3\times10^{-4}$---but this has no impact on the 
efficiency, provided a sufficient number of snapshots are available for reweighting. However, it is clear that 
$f$ will be greatly reduced in systems which undergo a folding transition upon 
lowering the temperature. This is simply a reflection of the fact that there is negligible overlap between the folded 
and unfolded distributions. In such cases, a useful reweighting protocol would require the generation of astronomical 
numbers of structures in the $T_M$ distribution, and annealing is advised. 

\section{Conclusion}
We have demonstrated the application of Neal's annealed importance sampling (AIS) algorithm for equilibrium 
sampling of the dileucine peptide. AIS 
allows the calculation of equilibrium averages from a nonequilibrium sample of strutures that results from a 
simulated annealing protocol. To our knowledge, AIS has not previously been applied to a molecular system. 
While the method, as na\"ively implemented here, represents only a modest improvement over  
constant temperature simulation, it is interesting for several reasons beyond equilibrium sampling. 

First, in applications where 
simulated annealing is already in widespread use (most notably, NMR structure 
calculations\cite{xplor-nih,wutrich-dyana-jmb97,brunger-cartSA-structure97}), the path weights 
may be used to calculate (perhaps noisy) equilibrium averages, and perhaps ultimately Boltzmann-distributed ensembles. 
The path weights also contain information 
that can be used to discriminate between different schedules, which may provide a way to optimize the schedule, 
based on the analysis of $t_{\text{ann}}$, the cost of annealing to ``good'' structures.

Second, it may be possible to improve considerably on the efficiency of the method by implementing a more 
sophisticated version, which uses a resampling procedure to prune the low weight paths at each cooling step. 
(For a detailed discussion of resampling methods, see the book by Liu\cite{liu-book}.) In this approach, we 
first cool some number $N$ of structures from the high temperature ($T_M$) ensemble, yielding $N$ weighted structures 
at $T_{M-1}$. We then resample $N$ times from this $T_{M-1}$ ensemble, according to the cumulative distribution 
function of the weights, pruning the low weight paths without biasing the sample. This type of 
approach was recently applied successfully to sampling near native protein configurations of a discretized and 
coarse-grained model\cite{liu-proteins07}. Nevertheless, we emphasize that the ultimate efficiency of any AIS 
protocol limited by the intrinsic sampling rate of the highest temperature, which may be modest; see Ref.\ \ref{repex-note}.

Finally, the AIS procedure could be naturally combined with ``annealing'' in the parameters of the Hamiltonian. 
Such a hybrid of AIS and Hamiltonian switching might be used, for example, to transform an NMR target 
function into a molecular mechanics 
potential function, over the course of a structure calculation. The result of such a calculation would be 
an equilibrium ensemble of structures, distributed according to the molecular mechanics potential. Such 
ensembles would find wide application, for instance in docking or homology modeling.

{\bf Acknowledgements}  The authors thank Gordon Rule for several enlightening discussions about NMR methodology. D.\ Z.\ 
thanks Chris Jarzynski for alerting him to Neal's work on AIS. This research was supported by the NSF (MCB-0643456), 
the NIH (GM076569), and the Department of Computational Biology, University of Pittsburgh.

\bibliographystyle{unsrt}
\bibliography{/home/elyman/latex/bib/my}
\clearpage

\begin{table}
\begin{center}
\begin{tabular}{|c|c|c|c|c|c|}
\hline \bf{T-steps} & \bf{Annealing time}& \bf{Successful}  & \bf{Fractional}    & \bf{Net cost}     & \bf{Efficiency} \\ 
		    & $t_{\text{anneal}}$ & \bf{structures}  & \bf{success rate}  &                  & \bf{gain}\\ 
		$M$ & $=(M-2)t_R$         & $n$         & $f\equiv n/N$ & $t_{\text{cost}}$ (nsec)   & $R$ \\
\hline  $3^{\dagger} $ & $0.5$  psec & $7.1$   & $4.4\times10^{-4}$ & $1.94$ & $1.5$  \\
\hline  $5^{\dagger} $ & $1.5$  psec & $ 43.7$ & $2.7\times10^{-3}$ & $1.36$ & $2.2$\\
\hline  $17^{\dagger}$ & $1.5$  psec & $137.6$ & $8.6\times10^{-3}$ & $0.97$ & $3.1$\\ 
\hline  $33$ & $1.5$  psec & $46.2$  & $2.9\times10^{-3}$ & $1.32$ & $2.3$\\ 
\hline  $9^{\dagger} $ & $3.5$  psec & $163.2$ & $1.0\times10^{-2}$ & $1.15$ & $2.6$\\
\hline  $17$ & $7.5$  psec & $205.3$ & $1.3\times10^{-2}$ & $1.38$ & $2.2$\\ 
\hline  $17$ & $15.0$ psec & $237.2$ & $1.5\times10^{-2}$ & $1.80$ & $1.7$\\ 
\hline  $17$ & $30.0$ psec & $353.8$ & $2.2\times10^{-2}$ & $2.16$ & $1.4$\\ 
\hline

\hline
\end{tabular}
\doublespacing
\caption{Comparison of the efficiency of AIS between several cooling schedules.  
$n$ is given by Eq.\ \ref{ndef}, $t_{\text{cost}}$ is given by Eq.\ \ref{aiscost}. The efficiency gain 
is the total simulation time invested in each successful annealed structure $t_{\text{cost}}$ divided 
by the time needed to generate an indepenendent structure by constant temperature simulation. The $^\dagger$ 
indicates schedules for which data are presented in Figs.\ \ref{ehistofig} and \ref{weighthistofig}.
}
\label{table1}
\end{center}
\end{table}

\clearpage
\section*{Figure Legends}
\subsubsection*{Figure~\ref{ehistofig}.}
Distribution of energies, from standard, constant temperature simulation and AIS. The dashed line
is the $T=500$ K distribution that was used for the high $T$ ensemble. The other data compare a $300$ nsec,
$T=298$ K constant temperature simulation to $298$ K ensembles generated by the AIS algorithm with different
cooling schedules. The schedules are discussed in Table \ref{table1}.

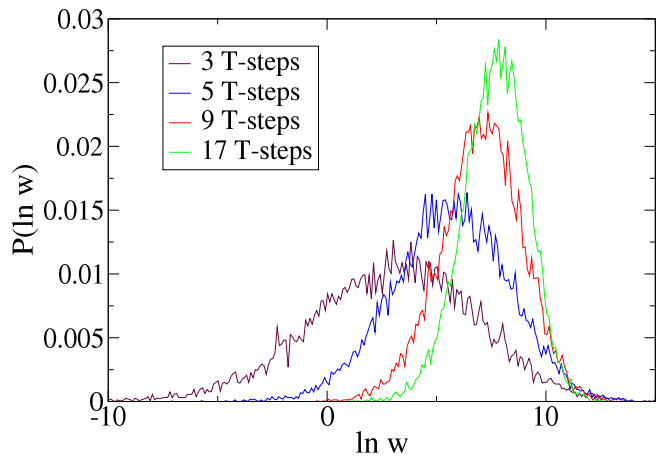
\begin{figure}
\includegraphics[width=8.7cm]{logw-histo.eps}
\caption{
}
\label{weighthistofig}
\end{figure}

\subsubsection*{Figure~\ref{weighthistofig}.}
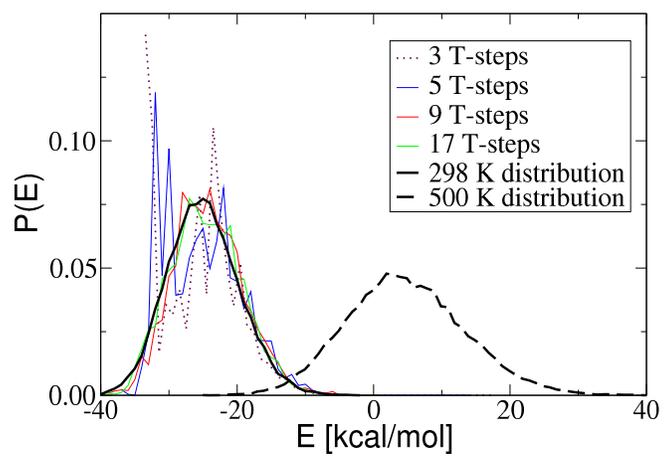
\begin{figure}
\includegraphics[width=8.7cm]{Ehistos.eps}
\caption{
}
\label{ehistofig}
\end{figure}

Distribution of the logarithm of trajectory weights for the four cooling schedules used
in Fig.\ \ref{ehistofig} and discussed in Table \ref{table1}.

\clearpage

\clearpage

\end{document}